\documentclass[structabstract]{aa}  
\usepackage{graphicx}
\usepackage{txfonts}
\usepackage{natbib}
\def\teff{\mbox{$T_{\rm eff}$}}
\def\ebv{\mbox{$E(4405-5495)$}}
\def\rv{\mbox{$R_{5495}$}}

\def\kl{\mbox{$k(\lambda)$}}
\begin{document}
   \title{UV Extinction Towards a Quiescent Molecular Cloud in the SMC}


   \author{J. Ma{\'\i}z Apell{\'a}niz\inst{1}
          \and
	  M. Rubio\inst{2}
          }

   \institute{Instituto de Astrof{\'\i}sica de Andaluc{\'\i}a-CSIC, Glorieta de la Astronom\'{\i}a s/n, E-18008 Granada, Spain \\
	      \email{jmaiz@iaa.es}
         \and
              Departamento de Astronom{\'\i}a, Universidad de Chile, Casilla 36-D, Santiago, Chile \\
              \email{monica@das.uchile.cl}
             }

   \titlerunning{UV Extinction Towards a Quiescent Molecular Cloud in the SMC}
   \authorrunning{J. Ma{\'\i}z Apell{\'a}niz \& M. Rubio}

   \date{Received xx xx 2011; accepted xx xx 2011}

 
  \abstract
   {The mean UV extinction law for the Small Magellanic Cloud (SMC) is usually taken as a template for low-metallicity galaxies. However, its current derivation is 
based on only five stars, thus placing doubts on its universality. An increase in the number of targets with measured extinction laws in the SMC is necessary to 
determine its possible dependence on parameters such as metallicity and star-forming activity.}
   {To measure the UV extinction law for several stars in the quiescent molecular cloud SMC B1-1.}
   {We obtained HST/STIS slitless UV spectroscopy of a 25\arcsec$\times$25\arcsec\ field of view and we combined it with ground-based NIR and visible photometry of the
stars in the field. The results were processed using the Bayesian photometric package CHORIZOS to derive the visible-NIR extinction values for each star. The
unextinguished Spectral Energy Distributions (SEDs) obtained in this way were then used to derive the UV extinction law for the four most extinguished stars. We also
recalculated the visible-NIR extinction for the five SMC stars with preexisting UV extinction laws.}
   {The UV extinction law for four SMC B1-1 stars within several pc of each other differs significantly from star to star. The 2175 \AA\ bump is moderately strong in 
one, weak in two, and absent in the fourth.}
   {}

   \keywords{Dust, extinction --- Galaxies: Magellanic Clouds --- Ultraviolet: ISM --- Stars: early-type}

   \maketitle
%

\section{Introduction}

	In the 1980s IUE was extensively used to derive the UV extinction law for a large number of stars in the Milky Way and the Magellanic Clouds. It became clear 
then that there were large variations among sightlines in the three galaxies. \citet{Cardetal89} used several tens of stars to derive a one-parameter family of
extinction laws for the Galaxy. The number of useful targets was even lower for the LMC and even more so for the SMC. The often quoted SMC extinction law of
\citet{Prevetal84} was based in only 3 stars (AzV 398, AzV 18/Sk 13, and Sk~191). Sk~191 turned out to be a poor choice since it is actually an unreddened object,
leaving only two useful objects. The reanalysis of \citet{GordClay98} added two further targets, AzV~214 and AzV~456/Sk~143, and \citet{Gordetal03} completed the current
list of SMC stars with a derived UV extinction law with a fifth target, AzV 23/Sk 17. 

	The characteristic that is most often quoted about the SMC extinction laws is the absence of the 2175 \AA\ wide absorption bump, which in the \citet{FitzMass90}
parameterization translates into a zero or very small $c_3$ term. However, this is not true of all SMC extinction laws since AzV 456/Sk 143 clearly shows the 
absorption feature.
The different location of AzV 456/Sk 143 in the SMC ``wing'' as compared with that of the other stars in the SMC bar prompted \citet{GordClay98} to compute two different SMC
extinction laws, one for the wing and one for the bar, based on one and three\footnote{As we have already mentioned, \citet{Gordetal03} added another star, AzV 23/Sk 17,
also located in the bar.} stars, respectively. The results of \citet{Gordetal03} also suggest another possible difference between the two regions: Their measured $R_V$
values for the four bar stars are in the typical range of 2.40-3.30 while that of the wing star is, 2.05 $\pm$ 0.17. 
The existence of such detected variations in a small sample is an obvious sign of the need for a better sample to quantify the prevalence of the different extinction
laws and to understand the underlying reason for such variations. Another reason why a better sample is needed is the relatively low values of $E(B-V)$ for the five stars, 
0-15-0.26 \citep{Gordetal03}, which make the measured $R_V$ values have large uncertainties.

	Why should we care about the SMC extinction law? Besides the importance that it has by itself, because the SMC is our best template for low-metallicity galaxies.
Indeed, it was assumed for some time that the absence of the 2175 \AA\ bump in the SMC extinction law was a metallicity effect but the detection of such a feature in 
Azv 456/Sk 143 seems to invalidate that hypothesis. Several explanations are possible, such as the existence of metallicity gradients in the ISM or the effect of
``star-formation activity'', a term that refers to the possible destruction of dust grains by UV radiation or shocks from massive stars. Until we have a better sample we
will likely not be able to decide which explanation is the most plausible one.

We selected as our target SMC B1-1, 
a cold molecular cloud with no signs of star formation located at the southern end of the SMC bar. 
It has a molecular mass of 2.4$\cdot$10$^4$ M$_\odot$ \citep{Rubietal04} as derived from the CO(1-0) emission line and a virial mass determination using observations 
done with the Swedish ESO Submilllimeter Telescope (SEST) at La Silla Observatory. Of the stars with existing UV extinction laws, the closest one to SMC B1-1 is 
AzV 18/Sk 13, located $\sim$14\arcmin\ away.


\section{Observations and data description}

\subsection{STIS data}

\begin{figure*}
\centering
\includegraphics[width=\linewidth]{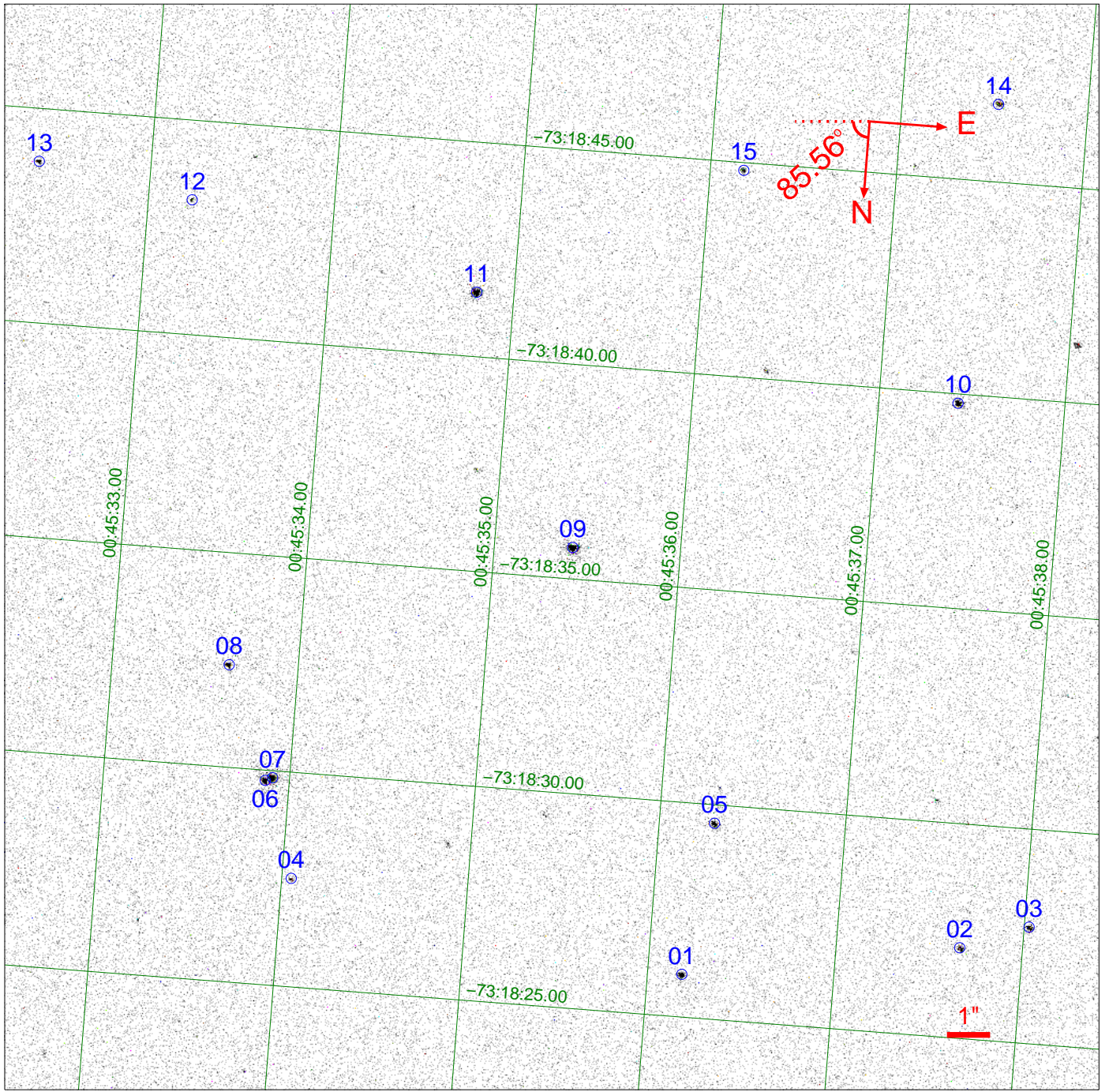}
\caption{F25CN270 STIS NUV-MAMA image of the SMC B1-1 region. The stars for which photometry was extracted are labelled.}
\label{nuv-image}
\end{figure*}

	SMC B1-1 was observed with the Space Telescope Imaging Spectrograph (STIS) aboard the Hubble Space Telescope (HST) on 27 May 2004. We
obtained four imaging exposures with STIS/NUV-MAMA, two imaging exposures with STIS/FUV-MAMA, and one slitless spectral exposure with
STIS/NUV-MAMA. Both MAMA detectors have no read noise and, in imaging mode, cover a field of 25\arcsec$\times$25\arcsec\ with 
1024$\times$1024 pixels. The NUV imaging exposures used two different filters, F25CN270 (centered at 2702 \AA, see Fig.~\ref{nuv-image}) and 
F25CN182 (centered at 1895 \AA) with total exposures times of 100 s for each filter. The FUV imaging exposures used a single filter, 
F25QTZ (centered at 1559 \AA) with a total exposure time of 200 s. The slitless spectral exposures, which cover an area slightly bigger than
that of the images, were obtained with the NUV objective prism (Fig.~\ref{nuv-prism}) with a total exposure time of 916 s. The NUV objective
prism has a variable dispersion that goes from $\approx$5~\AA/pixel at 1750 \AA\ to $\approx$45~\AA/pixel at 3000 \AA\ \citep{stis}. Our 
spectral exposures were obtained in the shadowed part of the orbit and with the F25SRF2 filter in order to minimize the effect of the 
geocoronal background.

	There are 15 stars that are easily detected in all three UV imaging filters and in the spectral exposures. They have been labelled in
Figs.~\ref{nuv-image}~and~\ref{nuv-prism} in order of ascending $y$ coordinate. Note that the stars have not been placed on an absolute reference frame,
so it is possible that the RA+dec grid is offset by $\sim$1\arcsec, as typical with coordinates derived from the header information in HST images. 

\subsection{NIR imaging}

	Deep NIR $JHK_{\rm s}$ imaging was obtained with ISAAC at the 8.2 m VLT telescope at Paranal Observatory on 24-25 September 1999 
(63.C-0329(A)). We used the short wavelength arm equipped with a $1024\times 1024$ 1024 pixels Hawaii Rockwell array, with a spatial resolution of 
0\farcs148\arcsec/pix and a total FOV of 2\farcm5$\times$2\farcm5.

	The observations were done in a series of 6 frames (NDIT), each individual frame with a 10 sec integration time (DIT) in each filter. The 
individual frames were coadded and a 60 sec image was stored. A 10 position dither mosaic was done with a separation of 15\arcsec\ using the same 
observing strategy to cover the SMC B1-1 region. In each filter, for every 10 minutes of on-source imaging we interleaved sky frame observations. 
These sky frames were chosen in a field with faint stars and no extended emission located 300\arcsec\ south  of the position of SMC B1-1. The sky 
field was observed in a similar way as the source frames. The procedure was repeated until we achieved a final integration time of 3600 s in $J$, 
$H$, and $K_{\rm s}$. 

	To produce the final images, each image was dark corrected, flat fielded, and sky subtracted, and then median averaged and combined using 
IRAF procedures. The final images were registered with respect to the $J$ image by means of several common stars. The final ISAAC/VLT images
cover a 2\farcm5$\times$5\farcm0 area and have limiting magnitudes of about $J=21$, $H=21$, and $K_{\rm s}=22$, respectively. 

	We identified the 15 stars observed in the HST field in the VLT/NIR images and performed aperture photometry using the IRAF/DAOPHOT package. 
We used the standard stars observed on both nights to calibrate the photometry in $K_{\rm s}$ (observed the first night) and in $H$ and $J$ 
(observed the second night). Additionally, we selected several stars from the 2MASS IR images with good photometry (AAA label) and compared their 
catalog magnitudes with our $JHK_{\rm s}$ photometry. The agreement between our photometry and 2MASS is good to 0.01 magnitudes. The NIR photometry 
in given in Table~\ref{nir_table}. Two of the stars (06 and 07) are too close together in the ground-based data to be resolved, so unresolved
magnitudes are provided.

\begin{table}
\caption{ISAAC/VLT NIR photometry.}
\label{nir_table} 
\centering        
\begin{tabular}{cccc} 
\hline\hline       
Star  & $J$                & $H$                & $K_{\rm s}$        \\
\hline                
01    & 19.414 $\pm$ 0.069 & 19.402 $\pm$ 0.072 & 19.525 $\pm$ 0.094 \\
02    & 20.002 $\pm$ 0.072 & 20.090 $\pm$ 0.074 & 19.968 $\pm$ 0.104 \\
03    & 20.086 $\pm$ 0.073 & 20.006 $\pm$ 0.076 & 19.963 $\pm$ 0.100 \\
04    & 19.864 $\pm$ 0.038 & 19.797 $\pm$ 0.045 & 19.647 $\pm$ 0.041 \\
05    & 18.836 $\pm$ 0.066 & 18.740 $\pm$ 0.068 & 18.718 $\pm$ 0.082 \\
06+07 & 17.987 $\pm$ 0.065 & 17.994 $\pm$ 0.068 & 18.000 $\pm$ 0.079 \\
08    & 18.436 $\pm$ 0.066 & 18.317 $\pm$ 0.068 & 18.235 $\pm$ 0.080 \\
09    & 18.424 $\pm$ 0.066 & 18.378 $\pm$ 0.068 & 18.307 $\pm$ 0.081 \\
10    & 18.812 $\pm$ 0.066 & 18.759 $\pm$ 0.068 & 18.739 $\pm$ 0.082 \\
11    & 18.464 $\pm$ 0.067 & 18.400 $\pm$ 0.069 & 18.426 $\pm$ 0.083 \\
12    & 20.246 $\pm$ 0.055 & 19.382 $\pm$ 0.035 & 18.877 $\pm$ 0.041 \\
13    & 19.402 $\pm$ 0.024 & 19.400 $\pm$ 0.034 & 19.258 $\pm$ 0.045 \\
14    & 17.658 $\pm$ 0.065 & 17.165 $\pm$ 0.067 & 17.066 $\pm$ 0.078 \\
15    & 17.236 $\pm$ 0.065 & 16.681 $\pm$ 0.067 & 16.566 $\pm$ 0.077 \\
\hline  
\end{tabular}
\end{table}

\subsection{Visible photometry}

	The Johnson-Cousins $UBVI$ photometry of the sources present in the STIS exposures was obtained from \citet{Zarietal02} using the 
VizieR \citep{Ochsetal00} J/AJ/123/855 catalog. 13 of the 15 stars are resolved in the \citet{Zarietal02} data while the pair 06+07 appears 
as a single unresolved source.

\begin{figure*}
\centering
\includegraphics[width=\linewidth]{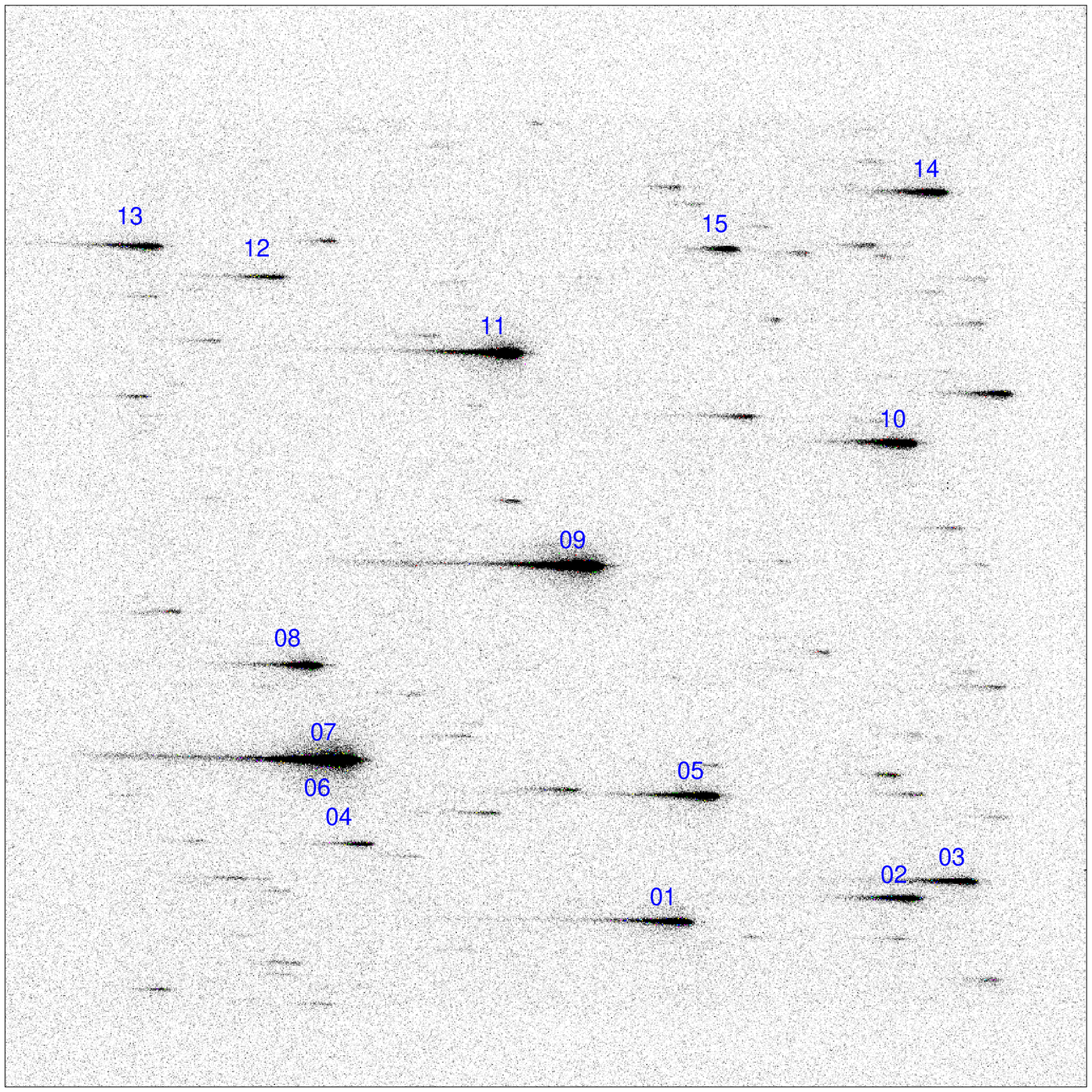}
\caption{STIS NUV-MAMA objective prism exposure of the SMC B1-1 region. See Fig.~\ref{nuv-image} for orientation and field size.}
\label{nuv-prism}
\end{figure*}


\section{Processing of the SMC B1-1 data}

	Our goal in this paper is to obtain the UV extinction law of as many of the stars in the SMC B1-1 region for which we can measure 
their spectra as possible. The strategy to accomplish that is:

\begin{enumerate}
  \item Measure the effective temperature (\teff) and optical/NIR extinction of the 14 sources which are point-like in the ground-based 
	  photometry.
  \item Extract the spectra for the 15 point sources in the NUV slitless exposures.
  \item Select the sources with the highest optical/NIR extinction, since those objects are the ones for which the UV extinction should be
	  more easily measured.
  \item Calculate the extinction law \kl\ for those stars selected in the previous step.
\end{enumerate}

\subsection{Effective temperatures and optical color excesses}

	The traditional mechanism to measure UV extinction laws is known as the pair method \citep{Massetal83} and requires obtaining spectra 
of two stars of the same spectral type, one with high extinction and the other one with low extinction. The fluxes of the two stars are 
corrected for distance effects if necessary, the low-extinction spectral energy distribution is divided by the high-extinction one, and the
result is normalized by the extinction difference between the two. The pair method requires the accurate measurement of the spectral type of the 
extinguished star, which may not be straightforward for dim objects, and the availability of data for a star with the same spectral type, 
which can lead to errors due to mismatches between stars.

	Alternatively, it is possible to measure extinction without referring to a standard star by using a synthetic spectral energy
distribution (SED) library and a numerical code that matches the available (spectro-)photometry to a grid with different stellar parameters
(e.g. \teff, $\log\, g$, and $Z$) and extinction amounts and laws \citep{Maiz04c,FitzMass05b}. In this way, it is possible to derive the intrinsic or
unextinguished SED directly from the data and, hence, calculate the extinction law directly by comparison with the observed spectral fluxes.
Here we will follow that approach for the optical/NIR
extinction using CHORIZOS \citep{Maiz04c,Maiz05d}, a Bayesian code that allows for the fitting of (spectro-)photometric data to an SED grid of 
up to five dimensions, including parameterized extinction laws\footnote{CHORIZOS is a public sofware written in IDL that can be downloaded 
from {\tt http://jmaiz.iaa.es}.}. For the amount of extinction CHORIZOS uses as parameter 
$\ebv \equiv A(4405) - A(5495)$, the monochromatic equivalent to $E(B-V)$, because the former depends only on the amount and type of 
dust while the latter also depends on the spectral type of the source\footnote{Throughout this work, wavelengths will be expressed in \AA\
unless otherwise explicitly stated.}. 
In other words, the same amount and type of dust in front of an O star and an M star will produce the same \ebv\ but different 
$E(B-V)$, so using $E(B-V)$ as an indicator of the amount of extinction can lead to some counterintuitive results (see e.g. 
\citealt{Massetal05a}) and should be avoided to characterize an extinction law.

	In order to use CHORIZOS to derive the intrinsic SED of the UV-bright objects in SMC B1-1 it is useful to take into account
different considerations. First of all, optical-NIR extinction laws have relatively simple functional forms and the Galactic laws can in 
principle be accurately described using a single-family parameter \citep{Cardetal89}\footnote{Recent works have determined that there may be more
variations than previously expected in the NIR extinction law but the effect of those variations should be relatively small in our case since,
as we will later see, our targets have all $A_K < 0.2$ magnitudes.}. In this wavelength range the published SMC extinction laws are 
not too different from the Galactic ones\footnote{The outstanding exception being that of AzV 23/Sk 17, but see below.}: for example, the SMC-bar law of 
\citet{GordClay98} is quite similar to the \citet{Cardetal89} law with \rv\ = 2.8 for $\lambda > 3000$ \AA\ ($\rv \equiv A(5495)/\ebv$ is the 
monochromatic equivalent to $R_V$). See also \citet{Maizetal12}.
In the UV, the extinction morphology is more complicated and, as \citet{FitzMass07} have shown, cannot be accurately described by a 
single-parameter family. Therefore, it makes sense to derive the intrinsic SEDs by excluding the photometric data with the shortest 
wavelengths because of the likely confusion between intrinsic (e.g. \teff) and extinction-law effects.
Thus, we will use the longest-wavelength UV photometric data (F25CN270) combined with the optical ($UBVI$) and NIR ($JHK$) data for that purpose. 
Second, given the location in the SMC and the observed magnitudes, we expect our stars to be early-to-mid B main-sequence 
stars with metallicity near $\log\, Z = -1.0$. Also, given that the optical-NIR colors of non-supergiant B stars are relatively insensitive
to gravity and metallicity, it will be appropriate to constrain our synthetic SED grid to main-sequence models with $\log Z\, = -1.0$ and to
use \teff, \ebv, and extinction law type as the free parameters in the grid. Note that the best photometric discriminant for the effective
temperature of B stars in the optical-NIR range is the Balmer jump and that with our filter selection we have two photometric points to its 
left (F25CN270 and $U$) and two points to its right ($B$ and $V$), thus allowing us to measure temperature precisely. 
Finally, in order to use consistent zero-points for the different
photometric systems involved, we will use the Johnson sensitivity curves as determined by \citet{Maiz06a} and the zero-point system of
\citet{Maiz07a}.

\begin{figure*}
\centering
\includegraphics[width=\linewidth, bb=0 170 595 742]{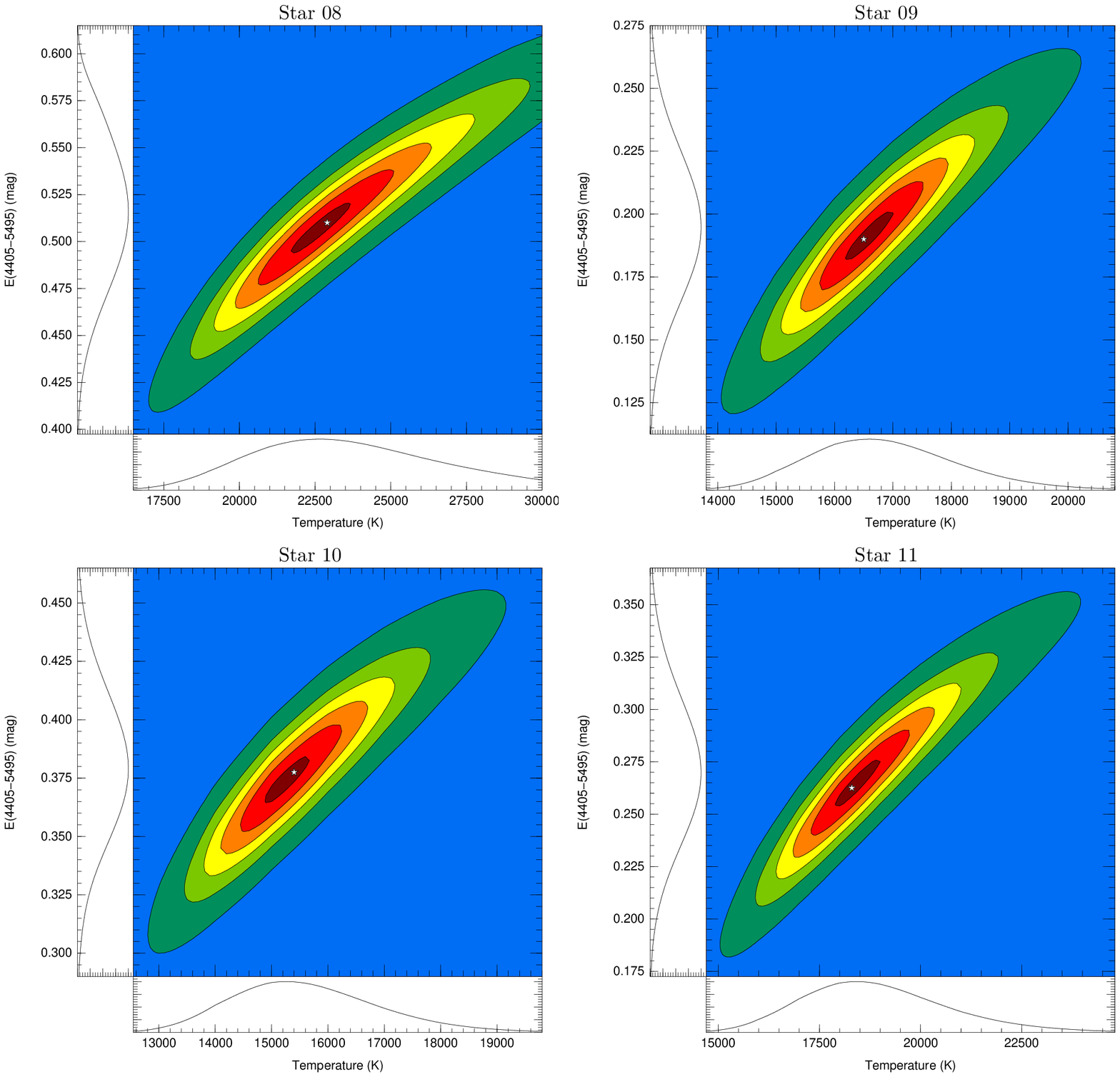}
\caption{Contour plots showing the likelihood in the \teff-\ebv\ plane for the four stars in the second CHORIZOS run with the highest mean
value for \ebv. The contour intervals are linearly spaced with the minimum and maximum at 0.05 and 0.95 times the peak (mode) value,
respectively. The star shows the location of the mode (not the mean) and the plots on each axis show the integrated likelihood for that 
parameter.}
\label{chorizos_cont}
\end{figure*}

\begin{figure*}
\centering
\includegraphics[width=\linewidth, bb=0 170 595 742]{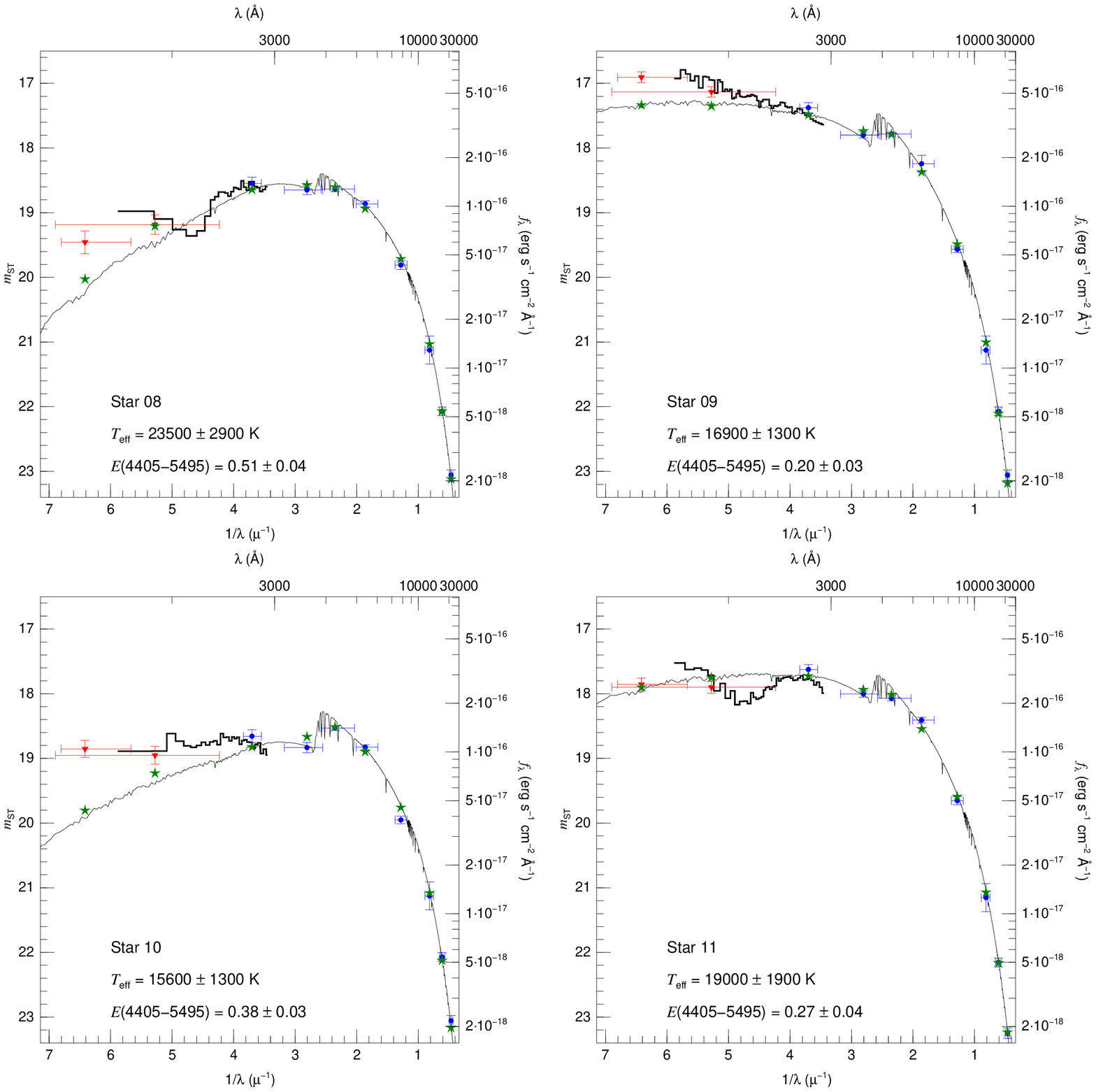}
\caption{MULTISPEC and fixed extinction-law CHORIZOS results for the four stars with the highest \ebv. The red and blue data points with 
error bars correspond to the observed UV-optical-NIR photometric magnitudes, with the vertical error bars showing the uncertainties and 
the horizontal ones the approximate extent of the filters. Only the blue data points points were used for the CHORIZOS fits. The 
continuous FUV-to-NIR line shows the CHORIZOS mode SED, which assumes a \citet{GordClay98} SMC-bar extinction law throughout the full 
wavelength range, while the NUV-only histogram shows the MULTISPEC extraction. The green stars 
show the synthetic photometry that corresponds to the CHORIZOS mode SED. The CHORIZOS values for \teff\ and \ebv\ are given for each star.}
\label{chorizos_spect}
\end{figure*}

\begin{table}
\caption{Second-run CHORIZOS results. Results for stars 14 and 15 are not shown due to the poor quality of the fits.}
\label{chorizos_table} 
\centering        
\begin{tabular}{cccc} 
\hline\hline       
Star  & \teff                   & \ebv            & reduced $\chi^2_{\rm min}$ \\
\hline                
01    & 13\,600 $\pm$    1300 K & 0.14 $\pm$ 0.05 & 1.1                        \\
02    & 12\,600 $\pm\;\;$ 900 K & 0.03 $\pm$ 0.03 & 6.8                        \\
03    & 12\,600 $\pm$    1100 K & 0.06 $\pm$ 0.04 & 1.8                        \\
04    & 13\,500 $\pm$    3800 K & 0.19 $\pm$ 0.15 & 0.2                        \\
05    & 11\,900 $\pm\;\;$ 800 K & 0.17 $\pm$ 0.03 & 3.7                        \\
06+07 & 13\,700 $\pm\;\;$ 700 K & 0.10 $\pm$ 0.03 & 3.8                        \\
08    & 23\,500 $\pm$    2900 K & 0.51 $\pm$ 0.04 & 1.5                        \\
09    & 16\,900 $\pm$    1300 K & 0.20 $\pm$ 0.03 & 2.1                        \\
10    & 15\,600 $\pm$    1300 K & 0.38 $\pm$ 0.03 & 4.8                        \\
11    & 19\,000 $\pm$    1900 K & 0.27 $\pm$ 0.04 & 2.8                        \\
12    & 12\,400 $\pm$    3100 K & 0.14 $\pm$ 0.04 & 3.2                        \\
13    & 10\,400 $\pm\;\;$ 800 K & 0.09 $\pm$ 0.06 & 2.2                        \\
\hline  
\end{tabular}
\end{table}

	Taking into account the considerations in the previous paragraph, we executed an initial CHORIZOS run for each of the 14 point-like
visible sources using as input the F25CN270 + $UBVI$ + $JHK$ photometry and as SEDs Kurucz atmospheres with main-sequence gravities and 
$\log\, Z = -1.0$ and \citet{Cardetal89} extinction laws, leaving three free parameters to fit (\teff, \ebv , and \rv). The results of that 
initial run show that all 14 sources have \ebv\ lower than 0.60 and only three have values above 0.25. Given the low extinction of the 11 
sources with $\ebv < 0.25$, their values of \rv\ are mostly unconstrained, as expected in such a case when the spectral types are not known a priori. 
For the three more extinguished stars we derive values of $\rv$ between 2.9 and 3.5 with uncertainties of $\approx$ 0.3 i.e. similar to those of 
the average Galactic and SMC extinction laws. This result prompted us to do
a second and final CHORIZOS run with the same photometric data and parameters but using a fixed extinction law, the SMC-bar of 
\citet{GordClay98}. The obtained values for \teff\ and \ebv\ in the final CHORIZOS run are similar to those of the first one but with slightly 
lower error bars. The results of the second run are shown in Table~\ref{chorizos_table} and plotted (for the four stars with the largest mean values 
of \ebv) in Figs.~\ref{chorizos_cont}~and~\ref{chorizos_spect}. As expected, all stars have temperatures 
consistent with being B stars\footnote{Very recently, \citet{Maizetal12} have derived a new optical/NIR extinction law for 30 Doradus.
Using that extinction law does not introduce fundamental changes to the results shown here.}. Their extinction-corrected luminosities are all 
consistent with being main-sequence stars.


\subsection{Spectral extraction}

	In order to extract the multiple spectra present in the objective prism data (Fig.~\ref{nuv-prism}), we used MULTISPEC 
\citep{Maiz05a}, a software package specifically designed for the extraction of multiple spectra from slitless HST exposures of crowded 
fields\footnote{MULTISPEC is a public sofware written in IDL that can be downloaded from {\tt http://jmaiz.iaa.es}.}.
MULTISPEC works by fitting multiple spatial profiles for each column in the spectral exposure. We used v2.0 of the code, which allows for the
use of tabulated spatial profiles and which has been applied by \citet{Knigetal08} to extract UV spectra from STIS/FUV-MAMA G140L
exposures of 47 Tuc. For the flux calibration of the data we used \citet{MaizBohl05} and included time-dependent sensitivity corrections.

	The extracted spectra for stars 08, 09, 10, and 11 are shown in Fig.~\ref{chorizos_spect} along with the mode CHORIZOS SED, which
assumes a \citet{GordClay98} SMC-bar extinction law. The extracted spectra have been binned in wavelength in order to obtain a uniform
S/N in each bin. The observed and model spectra have relatively similar values of $f_\nu$ in their 
common wavelength range, indicating that the extinction law for those stars cannot be too different from the SMC-bar one of \citet{GordClay98}. 
Note, however, that there are some readily apparent differences: stars 08 and 11 show a weak 2175 \AA\ absorption structure and the measured UV 
fluxes for stars 09 and 10 are larger than the model ones.

\subsection{Source selection}

	As previously mentioned only three point sources (08, 10, and 12) have $\ebv > 0.25$ and a fourth one (09) has a slightly lower value. 
Those will be the four sources for which we will calculate the UV extinction law since for the rest the relative uncertainties on \kl\ are
too large to yield a useful result. Note that previous studies have derived extinction laws for stars with even lower values of \ebv, but that
was possible because of the availability of data with higher S/N that what we are using here. 

\subsection{UV extinction-law calculation}

\begin{figure*}
\centering
\includegraphics[width=\linewidth, bb=0 170 595 742]{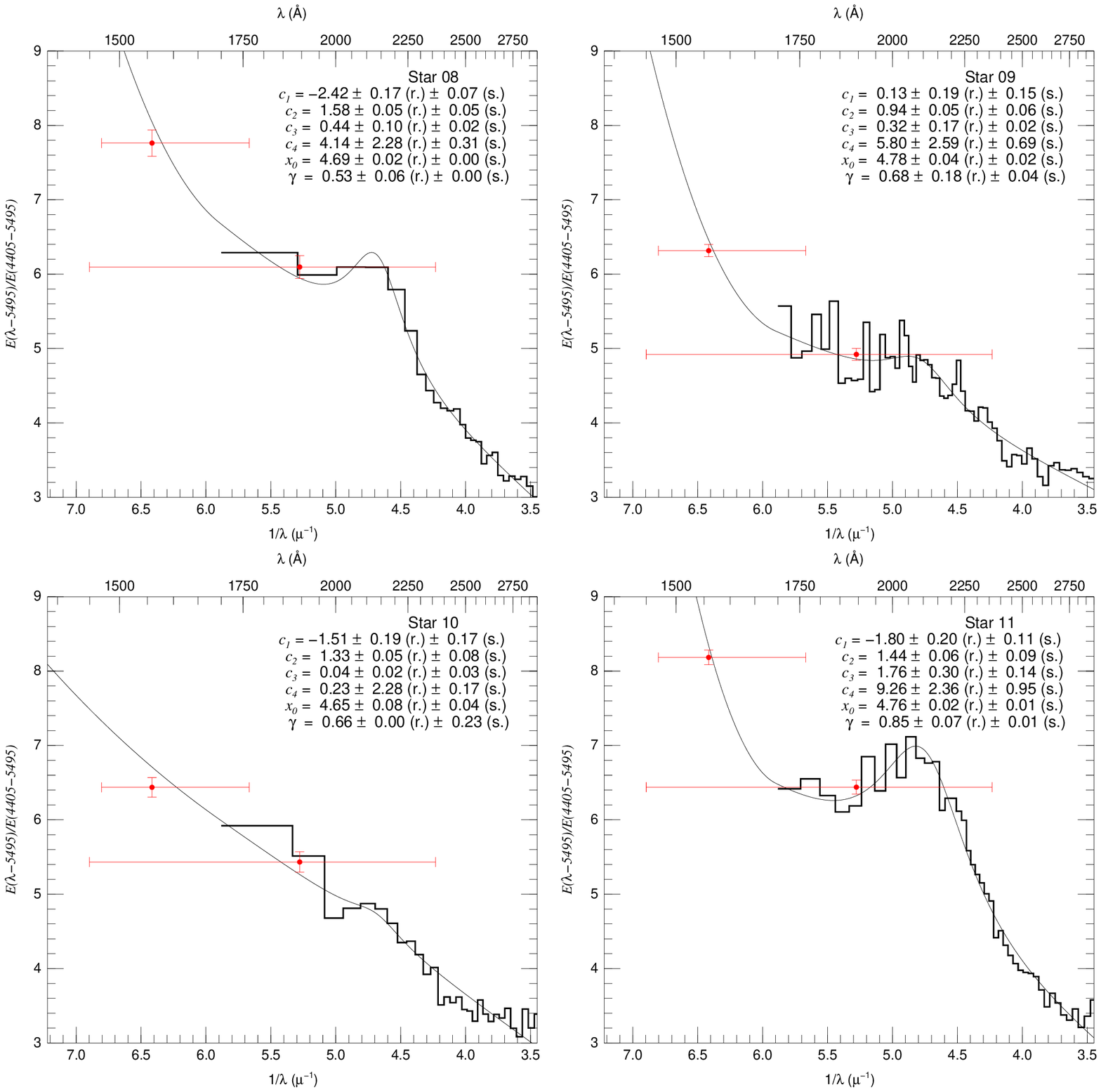}
\caption{Observed extinction laws (histogram used for the spectral data, points with error bars for the photometric data) and fitted 
\citet{FitzMass90} extinction laws (continuous line) for each of the four stars in Fig.~\ref{chorizos_spect}. The values of the fitted 
parameters are given in each case with their random and systematic errors indicated by ``r.'' and ``s.'', respectively.}
\label{extlaw}
\end{figure*}

	Finally, we obtain the UV extinction law \kl\ which, following the usual practice, we normalize as:

\begin{equation}
\kl = k(5495) + \frac{E(\lambda - 5495)}{\ebv}
\end{equation}

\noindent and then consider only the second term, which is 0 for $\lambda$~=~5495~\AA\ and 1 for $\lambda$~=~4405~\AA . Note, that, as 
previously explained, in this paper we use monochromatic rather than filter-integrated values for the definition of the extinction law. 
The mode value of $\kl - k(5495)$ is obtained by dividing the extracted spectrum for each star by the unextinguished version of the CHORIZOS-derived 
mode SED and normalizing the result. A
similar procedure can be done for the filter-integrated quantities derived from the measured F25QTZ and F25CN182. Both spectral and photometric
results are shown in Fig.~\ref{extlaw}. One readily apparent result is that there is a decrease in the intensity of the 2175 \AA\ absorption 
structure as we follow the stellar sequence 11 $\rightarrow$ 08 $\rightarrow$ 09 $\rightarrow$ 10. 

	One practical way of characterizing the UV extinction law is by using the six-parameter ($c_1$, $c_2$, $c_3$, $c_4$, $x_0$, $\gamma$) 
\citet{FitzMass90} recipe, which gives:

\begin{equation}
\kl - k(5495) = c_1 + c_2 x + c_3 D(x, x_0, \gamma) + c_4 F(x),
\end{equation}

\noindent where $x = 1/\lambda$ (usually expressed in $\mu^{-1}$),

\begin{equation}
D(x, x_0, \gamma) = \frac{x^2}{(x^2-x_0^2)^2+x^2\gamma^2},
\end{equation}

\noindent and

\begin{equation}
F(x) = \left\{ 
\begin{array}{ll}
0.5392(x-5.9)^2 + 0.05644(x-5.9)^3, & x > 5.9    \\
0,                                  & x \le 5.9.
\end{array}
\right.
\end{equation}

	We wrote a $\chi^2$-minimization code in order to fit such a \citet{FitzMass90} six-parameter function to the spectral (1700-2900 \AA) 
and photometric (F25QTZ and F25CN182) data for each of the four stars with higher extinction. When doing so, one has to be careful to account 
not only for random errors arising from the finite S/N of the data but also for systematic ones arising from possible spectral mismatches
(in our case, this corresponds to the uncertainty in the \teff\ and \ebv\ values derived by CHORIZOS). The results are shown and plotted 
in Fig.~\ref{extlaw}.

	As previously noted, there is a sequence of intensities of the 2175 \AA\ absorption structure (measured by $c_3$) in the order
11 $\rightarrow$ 08 $\rightarrow$ 09 $\rightarrow$ 10. It should be also pointed out that the values of the parameter that measure the
FUV slope ($c_4$) are anomalously high. However, those values are not very significative (their errors are very large) because they are
based mostly on a single photometric point (F25QTZ) centered at relatively long FUV wavelengths.

	Note that for our calculations we do not perform a prior subtraction of the Galactic contribution to the extinction. As described by 
\citet{Zagu07}, for the Magellanic Clouds that cannot be done from the data itself unless one assumes a smooth spatial distribution of the
Galactic cirri, something that the MIR observations do not support. Therefore, the UV extinction laws derived here are in principle a combination of a Galactic
and an SMC components. However, the existence of stars with very low extinction in the vicinity [e.g. star 02 has \ebv\ = 0.03 $\pm$ 0.03] argues in favor of the dominance 
of the SMC component. An additional point against the importance of a Galactic dust component is the wide range in reddenings in our sample, with values from 0.03 to 0.51 mag.
Note that at a distance of e.g. 1 kpc to a hypothetical Galactic cloud, 10\arcsec\ correspond to just 10\,000~A.U. To our knowledge, no Galactic cirrus shows such large
variations on such small scales. The low value of \ebv\ for star 02 is also in agreement with the foreground reddening of 0.037 measured by \citet{Schletal98} and with the
more recent work of \citet{SubrSubr10}, who find that most of the SMC has a Galactic foreground extinction lower than $A_V = 0.1$ mag.

\section{Looking back at the five SMC stars with preexisting extinction laws}

	In order to check that our methods are consistent with previous works, we used CHORIZOS to calculate the values of \ebv\ and \rv\ for the
five SMC stars with preexisting extinction laws. For that purpose, we used the \citet{Gordetal03} $UBVRI$ and the 2MASS $JHK_{\rm s}$ photometry
for those stars\footnote{Note that, being supergiants, they are much brighter than the SMC B1-1 stars. Hence, they all have 2MASS photometry with
small uncertainties.}. In this case we used a newer version of CHORIZOS that allows the stellar parameters to be \teff, luminosity class, and 
metallicity. By luminosity class, we mean a quantity that runs from 0.0 (hypergiants) to 5.5 (ZAMS) and, for a given \teff\ and metallicity,
yields a luminosity similar to the one derived from spectral classification (with Ia+ translated as 0.0 and Vz as 5.5). The SEDs in this case were 
TLUSTY \citep{LanzHube03,LanzHube07}. Three parameters were left fixed: \teff\ (from the spectral type), metallicity (SMC), and the logarithm of the
distance (4.782). Three parameters were allowed to vary: Luminosity class, \rv, and \ebv. For AzV 398 we excluded the $K_{\rm s}$ photometry from the
fit after an initial run because the star appears to have an excess in that band compared to the TLUSTY models, likely a wind effect. \citet{Cardetal89}
extinction laws were used. Results are
shown in Table~\ref{chorizos_table2}. Note that the values there for \rv\ and \ebv\ are monochromatic quantities while $A_V$ is filter-integrated. 

\begin{table*}
\caption{CHORIZOS results for the five stars with preexisting UV extinction laws.}
\label{chorizos_table2} 
\centering        
\begin{tabular}{lcccccc} 
\hline\hline       
\multicolumn{1}{c}{Star} & \teff     & luminosity class & \rv             & \ebv              & $A_V$             & red. $\chi^2_{\rm min}$ \\
\hline                
AzV 18/Sk 13             & 19\,000 K & 1.19 $\pm$ 0.01  & 3.93 $\pm$ 0.64 & 0.178 $\pm$ 0.022 & 0.696 $\pm$ 0.040 & 1.9                     \\
AzV 23/Sk 17             & 18\,000 K & 1.01 $\pm$ 0.01  & 3.41 $\pm$ 0.08 & 0.253 $\pm$ 0.003 & 0.878 $\pm$ 0.015 & 1.9                     \\
AzV 214                  & 23\,000 K & 1.80 $\pm$ 0.03  & 3.09 $\pm$ 0.20 & 0.246 $\pm$ 0.007 & 0.772 $\pm$ 0.034 & 2.2                     \\
AzV 398                  & 28\,000 K & 1.93 $\pm$ 0.03  & 3.57 $\pm$ 0.28 & 0.322 $\pm$ 0.018 & 1.162 $\pm$ 0.036 & 1.3                     \\
AzV 456/Sk 143           & 28\,000 K & 1.19 $\pm$ 0.02  & 2.63 $\pm$ 0.14 & 0.337 $\pm$ 0.011 & 0.901 $\pm$ 0.027 & 0.6                     \\
\hline  
\end{tabular}
\end{table*}

	The reduced $\chi^2_{\rm min}$ in all cases is quite good, indicating the good quality of the photometry. The (photometry-derived) 
luminosity classes are all between 1.0 and 2.0, as expected for supergiants. Our results for \ebv\ and \rv\ differ significantly from the 
band-integrated ones [$E(B-V)$ and $R_V$, respectively] of \citet{Gordetal03}. The explanation for the differences in \ebv\ is likely to be caused by the fact that we are measuring 
absolute extinctions while they are measuring extinctions relative to a comparison star. Indeed, our values for \ebv\ are larger than their values of $E(B-V)$
by 0.01 to 0.10 magnitudes, as expected under such circumstances. However, we also measure values of \rv\ that are systematically larger by $\sim$0.6. That is harder to assign 
to the inclusion of an additional \rv\ = 3.1 component in our values (the one responsible for the additional extinction), since that would pull them towards 3.1 and not increase
all of them independently of \rv. The difference may be attributed to the different methodologies and to the fact that our quantities are monochromatic and theirs
are band-integrated. 

	The uncertainties in Table~\ref{chorizos_table2} (and in Table~\ref{chorizos_table} as well) are formal (or random) ones. There are also 
systematic uncertainties caused by the mismatch between the SEDs used and the real ones and due to the problems with the \citet{Cardetal89} extinction laws
(fundamentally, the use of a seventh degree polynomial as a functional form in the optical range, see \citealt{Maizetal07,MaizSota08,Maizetal12}). 
However, an analysis of the photometric residuals indicates that the Balmer jump seems to be well fit. Furthermore, these 
objects have $\ebv < 0.5$, so the systematic effects of the \citet{Cardetal89} extinction laws should be small\footnote{A quite different situation compared to the
more extinguished case of Pismis 24-1 case discussed in \citet{Maizetal07}.}. Therefore, we do not expect the systematic uncertainties to be larger than the random 
ones, especially for \rv.

	The weighted mean of the \rv\ values for the four SMC bar stars is $3.36 \pm 0.07$ and the results are consistent with the four real \rv\ values being 
the same (three cases within one sigma, one between one and two sigma). Therefore, the results for those four stars and the three more extinguished stars in SMC
B1-1 are consistent with a visible-NIR extinction law not too different from the typical Galactic one, which has $\rv=3.1-3.2$. On the other hand, the result for
AzV~456/Sk~143 does indeed seem different, since its \rv\ is $\sim$5 sigmas lower. 


\section{Discussion}

	In this paper we have calculated for the first time the UV extinction law for four stars in SMC B1-1, a quiescent cloud in the Small Magellanic Cloud.
For those stars and for the other five stars with previously existing UV extinction laws we have also calculated their \ebv\ and \rv\ from visible-NIR photometry.

	For the four stars in SMC B1-1. all located within a few pc of each other, we have found significant variations in the extinction law and values. The latter is
simply a manifestation of the small-scale structure in the dust distribution while the former implies a different dust composition. The most clear variations in the UV 
extinction law are in the strength of the 2175 \AA\ bump, which goes from non-existing (as it is the case for the four of the SMC bar stars with preexisting laws) to a
case of moderate strength by Galactic standards though still weaker than the previously studied SMC bar star ($c_3 = 1.76 \pm 0.30$ for the new target, 
$2.57 \pm 0.22$ for the old one). $c_3$ does not appear positively correlated with \ebv\ (the most and the least extinguished stars have weak bumps while the two stars
with intermediate extinctions represent the extremes in bump strength) which argues against a foreground Galactic cloud as the source of the extinction. 

	Despite the significant UV variations, the SMC visible-NIR extinction law appears to be more uniform, at least within the current measurement errors. 
The variations in \rv\ have been attributed to different dust grain sizes for a long time \citep{BaadMink37}. All observed
eight SMC bar stars with well-measured \rv\ have values compatible with the typical Galactic one and only the SMC star measured in the wing has a somewhat lower \rv. However,
there are no clear cases of high values ($\rv \sim 5$), such as those found in some Galactic H\,{\sc ii} regions (see e.g. \citealt{Ariaetal06}). That is not surprising, given
that the Galactic cases with \rv\ are found within the bright areas of their H\,{\sc ii} regions and that is not the case for any of the analyzed SMC stars\footnote{AzV 214 is 
in NGC 346 but outside the central bright nebulosity.}.

	The existence of the 2175 \AA\ bump in stars located in relatively quiescent regions of the SMC argues in favor of its absence being caused by the presence of intense
UV radiation fields and stellar winds, which would destroy its carrier (see \citealt{Clayetal03,Sofietal06}). 
Nevertheless, our results also suggest that the final explanation may be more complex. Otherwise, the
four stars in SMC B1-1 would show the bump in their extinction laws. Also, since extinction appears to be patchy and associated with local dust clouds, $c_3$ should be
correlated with \ebv. Neither of those two effects appear in on our results.

\begin{acknowledgements}
This article is based on observations made with the NASA/ESA Hubble Space Telescope (HST) 
associated with GO program 9718 and obtained at the Space Telescope Science 
Institute, which is operated by the Association of Universities for Research in Astronomy, Inc., under NASA contract NAS 5-26555.
Support for both authors was provided by the Chilean FONDECYT grant number 1080335. J.M.A. also acknowledges support from
[a] the Spanish Government Ministerio de Educaci\'on y Ciencia through grants AYA2004-08260-C03, 
AYA2004-05395, AYA2007-64712, AYA2007-64052, and AYA2010-17631 and the Ram\'on y Cajal Fellowship program, co-financed with FEDER funds, and 
[b] the Consejer{\'\i}a de Educaci{\'o}n of the Junta de Andaluc{\'\i}a through grants TIC-101 and P08-TIC-4075. 
M.R. is also supported by the Chilean {\sl Center for Astrophysics} FONDAP grant number 15010003. The authors would like to thank the referee, Geoff Clayton, for his suggestions in
improving the paper, and Fran\c{c}ois Boulanger for his inspiration and support in the early stages of this project.
\end{acknowledgements}

\bibliographystyle{aa}
\bibliography{general}

\begin{thebibliography}{32}
\expandafter\ifx\csname natexlab\endcsname\relax\def\natexlab#1{#1}\fi

\bibitem[{Arias {et~al.}(2006)Arias, Barb{\'a}, Ma{\'\i}z~Apell{\'a}niz,
  Morrell, \& Rubio}]{Ariaetal06}
Arias, J.~I., Barb{\'a}, R.~H., Ma{\'\i}z~Apell{\'a}niz, J., Morrell, N.~I., \&
  Rubio, M. 2006, MNRAS, 366, 739

\bibitem[{Baade \& Minkowski(1937)}]{BaadMink37}
Baade, W. \& Minkowski, R. 1937, ApJ, 86, 123

\bibitem[{Cardelli {et~al.}(1989)Cardelli, Clayton, \& Mathis}]{Cardetal89}
Cardelli, J.~A., Clayton, G.~C., \& Mathis, J.~S. 1989, ApJ, 345, 245

\bibitem[{Clayton {et~al.}(2003)Clayton, Wolff, Sofia, Gordon, \&
  Misselt}]{Clayetal03}
Clayton, G.~C., Wolff, M.~J., Sofia, U.~J., Gordon, K.~D., \& Misselt, K.~A.
  2003, ApJ, 588, 871

\bibitem[{Fitzpatrick \& Massa(1990)}]{FitzMass90}
Fitzpatrick, E.~L. \& Massa, D. 1990, ApJS, 72, 163

\bibitem[{Fitzpatrick \& Massa(2005)}]{FitzMass05b}
Fitzpatrick, E.~L. \& Massa, D. 2005, AJ, 130, 1127

\bibitem[{Fitzpatrick \& Massa(2007)}]{FitzMass07}
Fitzpatrick, E.~L. \& Massa, D. 2007, ApJ, 663, 320

\bibitem[{Gordon \& Clayton(1998)}]{GordClay98}
Gordon, K.~D. \& Clayton, G.~C. 1998, ApJ, 500, 816

\bibitem[{Gordon {et~al.}(2003)Gordon, Clayton, Misselt, Landolt, \&
  Wolff}]{Gordetal03}
Gordon, K.~D., Clayton, G.~C., Misselt, K.~A., Landolt, A.~U., \& Wolff, M.~J.
  2003, ApJ, 594, 279

\bibitem[{Knigge {et~al.}(2008)Knigge, Dieball, Ma{\'{\i}}z~Apell{\'a}niz,
  Long, Zurek, \& Shara}]{Knigetal08}
Knigge, C., Dieball, A., Ma{\'{\i}}z~Apell{\'a}niz, J., {et~al.} 2008, ApJ,
  683, 1006

\bibitem[{Lanz \& Hubeny(2003)}]{LanzHube03}
Lanz, T. \& Hubeny, I. 2003, ApJS, 146, 417

\bibitem[{Lanz \& Hubeny(2007)}]{LanzHube07}
Lanz, T. \& Hubeny, I. 2007, ApJS, 169, 83

\bibitem[{Ma{\'\i}z~Apell{\'a}niz(2004)}]{Maiz04c}
Ma{\'\i}z~Apell{\'a}niz, J. 2004, PASP, 116, 859

\bibitem[{Ma{\'\i}z~Apell{\'a}niz(2005{\natexlab{a}})}]{Maiz05d}
Ma{\'\i}z~Apell{\'a}niz, J. 2005{\natexlab{a}}, in ESA Special Publication,
  Vol. 576, The Three-Dimensional Universe with Gaia, ed. C.~Turon, K.~S.
  O'Flaherty, \& M.~A.~C. Perryman, 449--+

\bibitem[{Ma{\'\i}z~Apell{\'a}niz(2005{\natexlab{b}})}]{Maiz05a}
Ma{\'\i}z~Apell{\'a}niz, J. 2005{\natexlab{b}}, STIS Instrument Science Report
  2005-02 (STScI: Baltimore)

\bibitem[{Ma{\'\i}z~Apell{\'a}niz(2006)}]{Maiz06a}
Ma{\'\i}z~Apell{\'a}niz, J. 2006, AJ, 131, 1184

\bibitem[{Ma{\'\i}z~Apell{\'a}niz(2007)}]{Maiz07a}
Ma{\'\i}z~Apell{\'a}niz, J. 2007, in ASP Conf. Series, Vol. 364, The Future of
  Photometric, Spectrophotometric and Polarimetric Standardization, ed.
  C.~Sterken, 227

\bibitem[{Ma{\'\i}z~Apell{\'a}niz \& Bohlin(2005)}]{MaizBohl05}
Ma{\'\i}z~Apell{\'a}niz, J. \& Bohlin, R.~C. 2005, STIS Instrument Science
  Report 2005-01 (STScI: Baltimore)

\bibitem[{Ma{\'{\i}}z~Apell{\'a}niz \& Sota(2008)}]{MaizSota08}
Ma{\'{\i}}z~Apell{\'a}niz, J. \& Sota, A. 2008, in Rev. Mex. Astron.
  Astrof{\'\i}s. (conference series), Vol.~33, 44--46

\bibitem[{Ma{\'\i}z~Apell{\'a}niz {et~al.}(2007)Ma{\'\i}z~Apell{\'a}niz,
  Walborn, Morrell, Niemel{\"a}, \& Nelan}]{Maizetal07}
Ma{\'\i}z~Apell{\'a}niz, J., Walborn, N.~R., Morrell, N.~I., Niemel{\"a},
  V.~S., \& Nelan, E.~P. 2007, ApJ, 660, 1480

\bibitem[{Ma{\'{\i}}z~Apell{\'a}niz {et~al.}(2012)}]{Maizetal12}
Ma{\'{\i}}z~Apell{\'a}niz, J. {et~al.} 2012, in preparation

\bibitem[{Massa {et~al.}(1983)Massa, Savage, \& Fitzpatrick}]{Massetal83}
Massa, D., Savage, B.~D., \& Fitzpatrick, E.~L. 1983, ApJ, 266, 662

\bibitem[{Massey {et~al.}(2005)Massey, Plez, Levesque, Olsen, Clayton, \&
  Josselin}]{Massetal05a}
Massey, P., Plez, B., Levesque, E.~M., {et~al.} 2005, ApJ, 634, 1286

\bibitem[{Ochsenbein {et~al.}(2000)Ochsenbein, Bauer, \& Marcout}]{Ochsetal00}
Ochsenbein, F., Bauer, P., \& Marcout, J. 2000, A\&AS, 143, 23

\bibitem[{Pr{\'e}vot {et~al.}(1984)Pr{\'e}vot, Lequeux, Maurice, Pr{\'e}vot, \&
  Rocca-Volmerange}]{Prevetal84}
Pr{\'e}vot, M.~L., Lequeux, J., Maurice, E., Pr{\'e}vot, L., \&
  Rocca-Volmerange, B. 1984, A\&A, 132, 389

\bibitem[{Rubio {et~al.}(2004)Rubio, Boulanger, Rantakyro, \&
  Contursi}]{Rubietal04}
Rubio, M., Boulanger, F., Rantakyro, F., \& Contursi, A. 2004, A\&A, 425, L1

\bibitem[{Schlegel {et~al.}(1998)Schlegel, Finkbeiner, \& Davis}]{Schletal98}
Schlegel, D.~J., Finkbeiner, D.~P., \& Davis, M. 1998, ApJ, 500, 525

\bibitem[{Sofia {et~al.}(2006)Sofia, Gordon, Clayton, Misselt, Wolff, Cox, \&
  Ehrenfreund}]{Sofietal06}
Sofia, U.~J., Gordon, K.~D., Clayton, G.~C., {et~al.} 2006, ApJ, 636, 753

\bibitem[{STScI(2007)}]{stis}
STScI. 2007, STIS Instrument Handbook, J. Kim Quijano, S. Hoifeltz, and J.
  Ma{\'\i}z Apell{\'a}niz (eds.)

\bibitem[{Subramaniam \& Subramanian(2010)}]{SubrSubr10}
Subramaniam, A. \& Subramanian, S. 2010, Interstellar Matter and Star
  Formation: A Multi-wavelength Perspective, ASI Conference Series, Edited by
  D.~K.~Ojha, 1, 107

\bibitem[{Zagury(2007)}]{Zagu07}
Zagury, F. 2007, \apss, 312, 113

\bibitem[{Zaritsky {et~al.}(2002)Zaritsky, Harris, Thompson, Grebel, \&
  Massey}]{Zarietal02}
Zaritsky, D., Harris, J., Thompson, I.~B., Grebel, E.~K., \& Massey, P. 2002,
  AJ, 123, 855

\end{thebibliography}

\end{document}